\newcommand {\BS}{{{\it Beppo}SAX}\ \ignorespaces}
\def\gta{ \lower .75ex \hbox{$\sim$} \llap{\raise .27ex \hbox{$>$}} }
\def\lta{ \lower .75ex\hbox{$\sim$} \llap{\raise .27ex \hbox{$<$}} }
\def \SAIT #1 #2 {{\em Mem.\ Soc.\ Astron.\ It.\/} {\bf #1}, #2}
\def \MESS #1 #2 {{\em The Messenger\/} {\bf #1}, #2}
\def \ASTRNACH #1 #2 {{\em Astron. Nach.\/} {\bf #1}, #2}
\def \AAP #1 #2 {{\em Astron. Astrophys.\/} {\bf #1}, #2}
\def \AAL #1 #2 {{\em Astron. Astrophys. Lett.\/} {\bf #1}, L#2}
\def \AAR #1 #2 {{\em Astron. Astrophys. Rev.\/} {\bf #1}, #2}
\def \AAS #1 #2 {{\em Astron. Astrophys. Suppl. Ser.\/} {\bf #1}, #2}
\def \AJ #1 #2 {{\em Astron. J.\/} {\bf #1}, #2}
\def \ANNREV #1 #2 {{\em Ann. Rev. Astron. Astrophys.\/} {\bf #1}, #2}
\def \APJ #1 #2 {{\em Astrophys. J.\/} {\bf #1}, #2}
\def \APJL #1 #2 {{\em Astrophys. J. Lett.\/} {\bf #1}, L#2}
\def \APJS #1 #2 {{\em Astrophys. J. Suppl.\/} {\bf #1}, #2}
\def \APSS #1 #2 {{\em Astrophys. Space Sci.\/} {\bf #1}, #2}
\def \ASR #1 #2 {{\em Adv. Space Res.\/} {\bf #1}, #2}
\def \BAIC #1 #2 {{\em Bull. Astron. Inst. Czechosl.\/} {\bf #1}, #2}
\def \JSQRT #1 #2 {{\em J. Quant. Spectrosc. Radiat. Transfer\/} {\bf #1}, #2}
\def \MN #1 #2 {{\em Mon. Not. R. Astr. Soc.\/} {\bf #1}, #2}
\def \MEM #1 #2 {{\em Mem. R. Astr. Soc.\/} {\bf #1}, #2}
\def \PLR #1 #2 {{\em Phys. Lett. Rev.\/} {\bf #1}, #2}
\def \PASJ #1 #2 {{\em Publ. Astron. Soc. Japan\/} {\bf #1}, #2}
\def \PASP #1 #2 {{\em Publ. Astr. Soc. Pacific\/} {\bf #1}, #2}
\def \NAT #1 #2 {{\em Nature\/} {\bf #1}, #2}

\documentstyle{memsait}
\input epsf.sty
\begin{opening}
\title{A SELF-CONSISTENT TEST OF COMPTONISATION MODELS USING BEPPOSAX
OBSERVATIONS OF SEYFERT GALAXIES} 
\author{P.O. Petrucci $^1$, F. Haardt$^2$, L. Maraschi$^1$,
P. Grandi$^3$, G. Matt$^6$, F. Nicastro$^{3,4,5}$, L. Piro$^3$,
G.C. Perola$^6$, A. De Rosa$^3$}
\institute{$^1$ Osservatorio Astronomico di Brera, Milano, Italy, $^2$
Universit\'a dell'Insubria, Como, Italy, $^3$ IAS/CNR, Roma, Italy, $^4$ CfA,
Cambridge Ma., USA, $^5$ Osservatorio Astronomico di Roma, Roma, Italy, $^6$
Universit\'a degli Studi ``Roma 3'', Roma, Italy}
\date{} 
\end{opening}

\begin{document}

\oddpagefooter{}{}{} 
\evenpagefooter{}{}{} 

\begin{abstract}
We test accurate models of Comptonisation spectra over BeppoSAX
observations of Seyfert galaxies focusing on the long look at NGC 5548.
The hot plasma temperature derived with these models is significantly
higher than that obtained fitting the same data with a power law plus
high energy cut off model for the continuum. This is due to the fact that
in anisotropic geometries Comptonisation spectra show "intrinsic"
curvature which moves the fitted high energy cut-off to higher
energies. We also show preliminary results of our analysis of a sample of
5 other Seyfert galaxies.
\end{abstract}

\section{The Comptonisation model: the anisotropy break}
\begin{figure}[b]
\vspace*{-0.5cm}
\epsfxsize=7cm
\begin{tabular}{ll}
\begin{minipage}{7cm}
\epsfbox{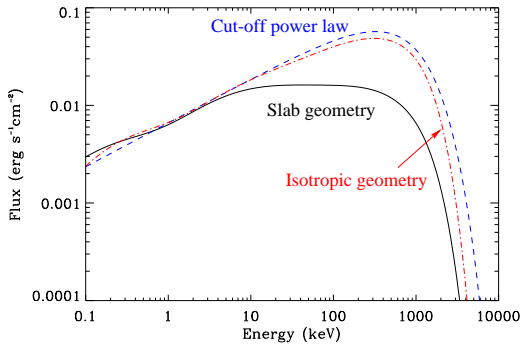}
\end{minipage} &
\begin{minipage}{6cm}
\caption[h]{Comptonised models for isotropic and anisotropic (slab)
geometries assuming $kT_e=360$ keV. In each case, the optical depths have
been chosen so as to produce approximatively the same spectral index in
the 2-10 keV X--ray range. We have also over-plotted a cut--off power law
with $E_{\rm c}=2kT_{\rm e}$ in dashed line.}
\end{minipage}
\end{tabular}
\vspace*{-0.5cm}
\end{figure}
Soft photons emitted by a flat optically thick surface, comptonised in a
hot corona with plane parallel (slab) geometry, emerge with anisotropic
spectra due to the asymmetry of the forward/backward scattering in the
first scattering orders. The spectrum observed at 30$^\circ$ inclination
from the normal to the plane (computed with the codes of Haardt, 1994 and
Poutanen and Svensson, 1996) is shown in Fig. 1. It can be approximately
described by a broken power law, the energy of the break $E_{\rm break}$
roughly lying between the second and the third scattering order peaks
(Haardt, 1993). For comparison we show on the same figure the spectrum
produced by comptonisation of an isotropic source of soft photons and the
power law + exponential cut-off approximation to the latter. We suppose a
corona temperature of 360 keV. The cut-off energy for the power law
approximation has been set to $E_c=2kT_e=720 keV$. Clearly there is a
large intrinsic difference in these continua which results in different
physical parameters when a given set of data is fitted with either
model. In the following we report the results of our study of NGC 5548
(Petrucci et al., 2000, hereafter P00) and preliminary results of other
objects.

\section{The long look at NGC 5548}
The average spectrum is well represented by a plane parallel corona with
an inclination angle of 30$^{\circ}$, a soft photon temperature of 5 eV
and a hot plasma temperature and optical depth of $kT_{\rm e}\simeq$ 360
keV and $\tau\simeq$ 0.1, respectively (P00). If energy balance applies,
such values suggest that, for the slab geometry, the hot gas is {\it
photon starved}, i.e., it is undercooled.

We show in Fig. 2a the best fit derived using Comptonisation (in slab
geometry) and a cut--off power law models. The (harder) power law model
requires a largely lower cut--off energy than that required by the
(intrinsically curved) slab Comptonisation spectrum and predicts a corona
temperature $kT_e\simeq 60$ keV. The two models require a different
normalization for the reflection component (larger for the slab) and are
roughly in agreement below $ 200$ keV, the upper energy end of our
data. However they differ by up to a factor 10 near 500 keV, due to their
different cut--off energy.

The source exhibits a flare during the central part of the observation.
The low--to--high state transition clearly indicates a change of the
Compton parameter $y$, i.e., of the Comptonised--to--soft luminosity
ratio (cf. Fig. 2b). It seems to be most probably due to an increase of
the cooling rate, rather than to a decrease of the heating rate, since we
observe a pivoting at high energies of the continuum in the two states
(cf. P00). If this interpretation is correct, then the spectral softening
in the high state is very naturally explained by a drop of the corona
temperature, ultimately due to an increase of the UV--EUV soft photon
flux (cf. P00).
\begin{figure}
\vspace*{-0.5cm}
\epsfxsize=13cm
\epsfbox{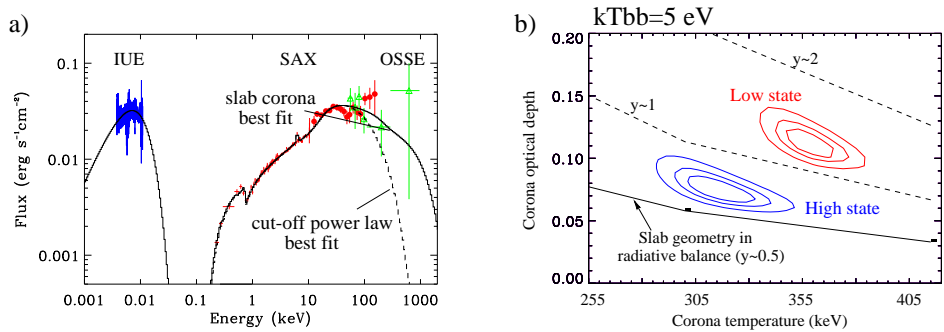}
\vspace*{-0.5cm}
\caption[h]{{\bf (a):} \BS data set of NGC 5548 with non--simultaneous
IUE and OSSE data (from Magdziark et al., 1998) with the corresponding
best fits Comptonisation model (in slab geometry, solid line) and simple
cut--off power law model (dashed line).{\bf (b):} Contour plots of $\tau$
vs. $\Theta$ in the low and high states for a slab corona configuration
for $kT_{bb}=5$ eV. We have also over-plotted the $\Theta-\tau$ relations
predicted when energy balance is achieved with the corresponding (rougly
constant) value of the Compton parameter (from Stern et al. 1995).}
\vspace*{-0.5cm}
\end{figure}

\section{Application to other Seyfert observations: preliminary results}
We apply the same type of analysis done for NGC 5548 to \BS observations
of other Seyfert galaxies. The best fit parameters using slab corona and
cut-off power law models are plotted in Fig. 3. We can see that: {\bf 1)}
the spectra of the different sources are consistent with a corona model
in radiative balance (Fig. 3a), {\bf 2)} the cut-off power law model
always under-estimates the corona temperature in comparison to the
Comptonisation model, up to a factor 6 for the observation of NGC 4151 in
1999 (Fig. 3b). {\bf 3)} we found a correlation between the corona
temperature, obtained using a cut-off power law model, and the photon
index (Fig 3c) whereas the two parameters are anticorrelated using a
\begin{figure}[h]
\vspace*{-0.5cm}
\caption[h]{}
\epsfxsize=13cm
\epsfbox{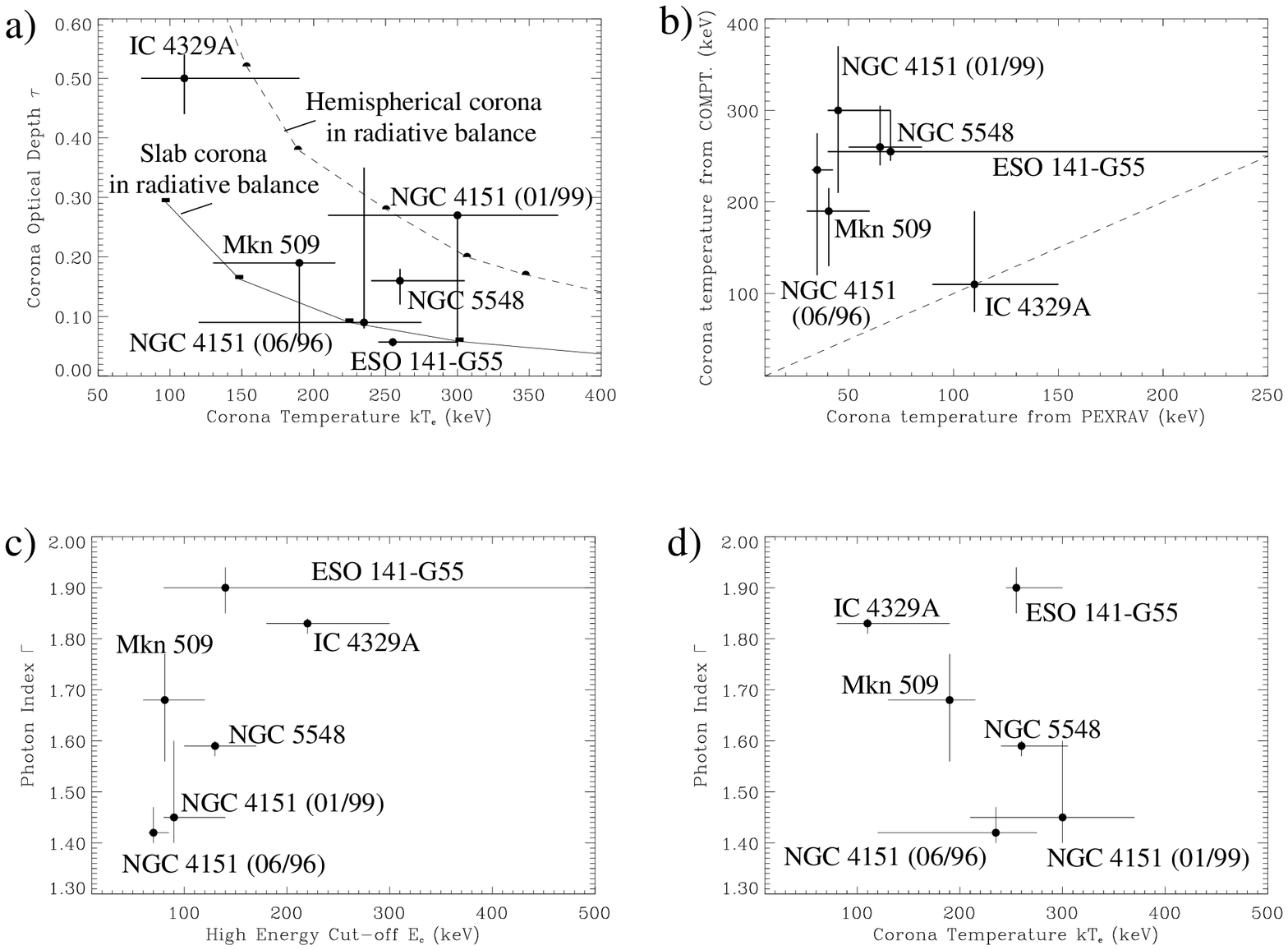}
\end{figure}
Comptonised slab corona model (Fig. 3d).

\vspace*{-1.cm}
\acknowledgements We gratefully acknowledge J. Poutanen for providing us
his code. Work supported by the European Commission under contract number
ERBFMRX-CT98-0195.


\end{document}